\documentclass[review]{elsarticle}

\usepackage{lineno,hyperref}
\modulolinenumbers[5]

\journal{Journal of \LaTeX\ Templates}

\begin{document}

\begin{frontmatter}
\title{Potential, Challenges and Future Directions for Deep Learning in Prognostics and Health Management Applications}
%
%
%
%
\author{Olga Fink, 
        Qin Wang}
\address{Intelligent Maintenance Systems, ETH Zurich}
\author{Markus Svens\'{e}n}
\address{GE Aviation Digital}
\author{Pierre Dersin}
\address{Alstom}
\author{Wan-Jui Lee}
\address{Dutch Railways}
\author{Melanie Ducoffe}
\address{Airbus}

%




\begin{abstract}
Deep learning applications have been thriving over the last decade in many different domains, including computer vision and natural language understanding. The drivers for the vibrant development of deep learning have been the availability of abundant data, breakthroughs of algorithms and the  advancements in hardware. Despite the fact that complex industrial assets have been extensively monitored and large amounts of condition monitoring signals have been collected, the application of deep learning approaches for detecting, diagnosing and predicting  faults of complex industrial assets has been limited. The current paper provides a thorough evaluation of the current developments, drivers, challenges, potential solutions and future research needs in the field of deep learning  applied to Prognostics and Health Management (PHM)  applications.
\end{abstract}


\end{frontmatter}
\section{Today's Challenges in PHM Applications}
The goal of Prognostics and Health Management (PHM) is to provide methods and tools to design optimal maintenance policies for a specific asset under its distinct operating and degradation conditions, achieving a high availability at minimal costs. 
PHM can be considered as a holistic approach to an effective and efficient system health management \cite{Lee2014}. PHM integrates the detection of an incipient fault (fault detection), its isolation, the identification of its origin and the specific fault type (fault diagnostics) and the prediction of the remaining useful life (prognostics). However, PHM does not end with the prediction of the remaining useful life (RUL). The system health management goes beyond the predictions of failure times and supports optimal maintenance and logistics decisions by considering the available resources, the operating context and the economic consequences of different faults. Health management is the process of taking timely and optimal maintenance actions based on outputs from diagnostics and prognostics, available resources and operational demand. PHM focuses on assessing and minimizing the operational impact of failures, and controlling maintenance costs \cite{Lee2014}.

Nowadays, the condition of complex systems is typically monitored by a large number of different types of sensors, capturing e.g. temperature, pressure, flow, vibration, images or even video streams of system conditions, resulting in very heterogeneous condition monitoring data at different time scales. Additionally, the signals are affected by measurement and transmission noise. In many cases, the sensors are partly redundant, having several sensors measuring the same system parameter. Not all of the signals contain information on a specific fault type since different fault types are affecting different signals and the correspondence is generally not one-to-one.

In most cases, using raw condition monitoring data in machine learning applications will not be conducive to detect faults or predict impending failures. Hence, successful PHM applications typically require manual pre-processing to derive more useful representations of signals in the data, a process known as \emph{feature engineering}. Feature engineering involves combinations of transforming raw data using e.g.\  statistical indicators or other signal processing approaches, such as time-frequency analysis, and steps to reduce the dimensionality of the data, where applicable, either with manual or automatic feature selection (filter, wrapper or embedded approaches) ~\cite{Forman2003}. 

Feature selection depends on the past observations or knowledge of the possible degradation types and their effect on the signals. Selecting too few or too many features may result in missed alarms, particularly for those fault types that have not been previously observed. At the same time, the number of false alarms must be minimized since this would otherwise impact the credibility of the developed model negatively. Thus, feature selection in a supervised fault classification problem must contribute to minimizing the false alarm rate (false positives) and maximizing the detection rate (true positives). 

The concept of feature extraction is also closely linked to the concept of condition indicators~\cite{sharma2016review}. A condition indicator is defined as a feature of condition monitoring system data whose behavior changes in a predictable way as the system degrades or operates in different operational modes. Therefore, a condition indicator can be any feature that enables to distinguish normal from faulty conditions or for predicting the remaining useful life (RUL). While there can be several condition indicators for one system, a more attractive way of monitoring the system health condition is by integrating several condition indicators into one health indicator, a value that represents the health status of the component to the end user. Health indicators have to follow some desired characteristics, including monotonicity, robustness and adaptability \cite{hu2016deep}. Several ways have been proposed to design health indicators and partly even subsequently using them for predicting RUL\cite{guo2017recurrent}. Recently, also an approach for learning the healthy system condition and using the distance measure to the learned healthy condition as an unbounded health indicator was proposed \cite{michau2020feature, arias2019knowledge}.

Due to the multiplicity of the possible fault types that can occur, feature engineering may face limitations to design a set of representative features that is able to depict the differences between all the possible fault types. Handcrafted features do not necessarily offer generalization ability and transferability from one system to another or even to other fault types. They also have a limited scalability due to the expert-driven manual approach. Additionally, the performance of feature engineering is highly dependent on the experience and expertise of the domain experts performing the task. The quality of the extracted features highly influences the performance of machine learning approaches that are using the extracted features. As the number of monitored parameters increases so does the difficulty of feature engineering for diagnostics engineers and consequently there is an interest in automating this processes~\cite{Yan2015} or circumventing the need for feature engineering altogether. Deep learning (DL) has the potential to incorporate feature engineering, or at least parts thereof, into the end-to-end learning processes.

While fault detection and fault diagnostics have been recently adopting DL approaches, prognostics has remained a rather difficult terrain for DL. Prognostics is the study and prediction of the future evolution of the health of the system being monitored, through the estimation of the RUL. Several directions have been proposed for estimating the RUL which broadly fall into the following categories ~\cite{lee2014prognostics, fink2020data}:
(a)model-based approaches (partly also referred to as physics-based), i.e. relying  mainly on multi-physics models for asset normal operation and physical degradation laws (possibly modelled as stochastic processes); (b)data-driven approaches that are based on condition monitoring data (where DL approaches belong to); (c)knowledge-based approaches, relying on expert judgements. There are also combinations of these three directions that are typically referred to as "hybrid approaches". While any combinations are in fact possible, the term "hybrid approaches" is typically used for approaches combining data-driven techniques with physics-based approaches ~\cite{arias19_hybrid}. 

Two different scenarios can be considered for the RUL prediction:
\begin{itemize}
  \item Degradation prediction (immediate degradation onset, dependent on the operating conditions)
 \item Detection of fault onset (e.g. crack initiation) and a subsequent prediction of the fault progression (dependent on the operating conditions)
\end{itemize}

Earlier works in data-driven RUL prediction have been mainly focusing on the statistical models, such as the thorough mathematical treatment of the RUL estimation in ~\cite{banjevic2009remaining}, with statistical  properties of the expectation and standard deviation of the RUL and their asymptotic behaviour, under broad assumptions. A fairly comprehensive survey of data-driven approaches for predicting RUL can be found in Si et al~\cite{si2011remaining}. A frequently used approach for data-driven RUL prediction consists of modelling degradation trajectories by stochastic processes that capture more or less accurately degradation physics. An  example of that approach is given by Zhai and Ye in~\cite{ZhaiRUL2017}. One of the research directions is also on assessing the performance of prognostics algorithms and the uncertainty in predictions by defining key performance indicators, such as the work of Saxena et al.~\cite{Saxena2014}.  A recent contribution has highlighted the importance of changes in the coefficient of variation of extracted features in detecting faults and in segmenting time series for health assessment and prognostics ~\cite{atamuradov2017}. This approach can be also linked to learning or extracting the health indicators and using them to detect the fault onset.



Ultimately, what PHM may expect from deep learning is either solving complex PHM problems that were not solvable with traditional approaches or improving the performance of the traditional approaches, automating the development of the applied models and making the usage of the models more robust and more cost-effective in industrial contexts.


Given its promising potential for PHM applications, DL is particularly well positioned to offer solutions to following issues:

\begin{itemize}
  \item Ability to automatically process massive amounts of condition monitoring data
  \item Ability to automatically extract useful features from high-dimensional, heterogeneous data sources
   \item Ability to learn functional and temporal relationships between and within the time series of condition monitoring signals
  \item Ability to transfer knowledge between different operating conditions and different units.
 
\end{itemize}

In addition, there are several general problems that need to be overcome in order to win broader acceptance of machine learning models in general and DL models in particular,  by the different stakeholders. These challenges include:

\begin{itemize}
  \item Ability to identify and deal with outliers
  \item Dynamic learning ability of the applied approaches: in particular, learning evolving operating conditions and distinguishing them from evolving faults 
  \item Ability to detect novelty, i.e.  hitherto unknown degradation types
  
  \item Robustness of the applied approaches, including cases under highly varying operating conditions
  \item Generalization ability of the developed models (ability to transfer knowledge between different operating conditions and different units)
  \item Interpretability of the results obtained for the domain experts
  \item Automated and optimal decision support taking the system condition and the relevant constraints, such as resource availability into consideration.
\end{itemize}

Currently, the application of DL in PHM is mainly driven by the developments in other DL application domains, such as  computer vision and natural language processing (NLP). Recent progress in DL has also been increasingly reflected in DL applications in PHM, although the transfer to industrial applications has been limited.  While some of the requirements of PHM are similar to those encountered in other fields, other requirements become more specific due to the specificities of complex industrial systems and the collected condition monitoring data. 

\begin{figure}[h]
\centering
\caption{Topics covered by this article.}
\includegraphics[width=0.9\textwidth]{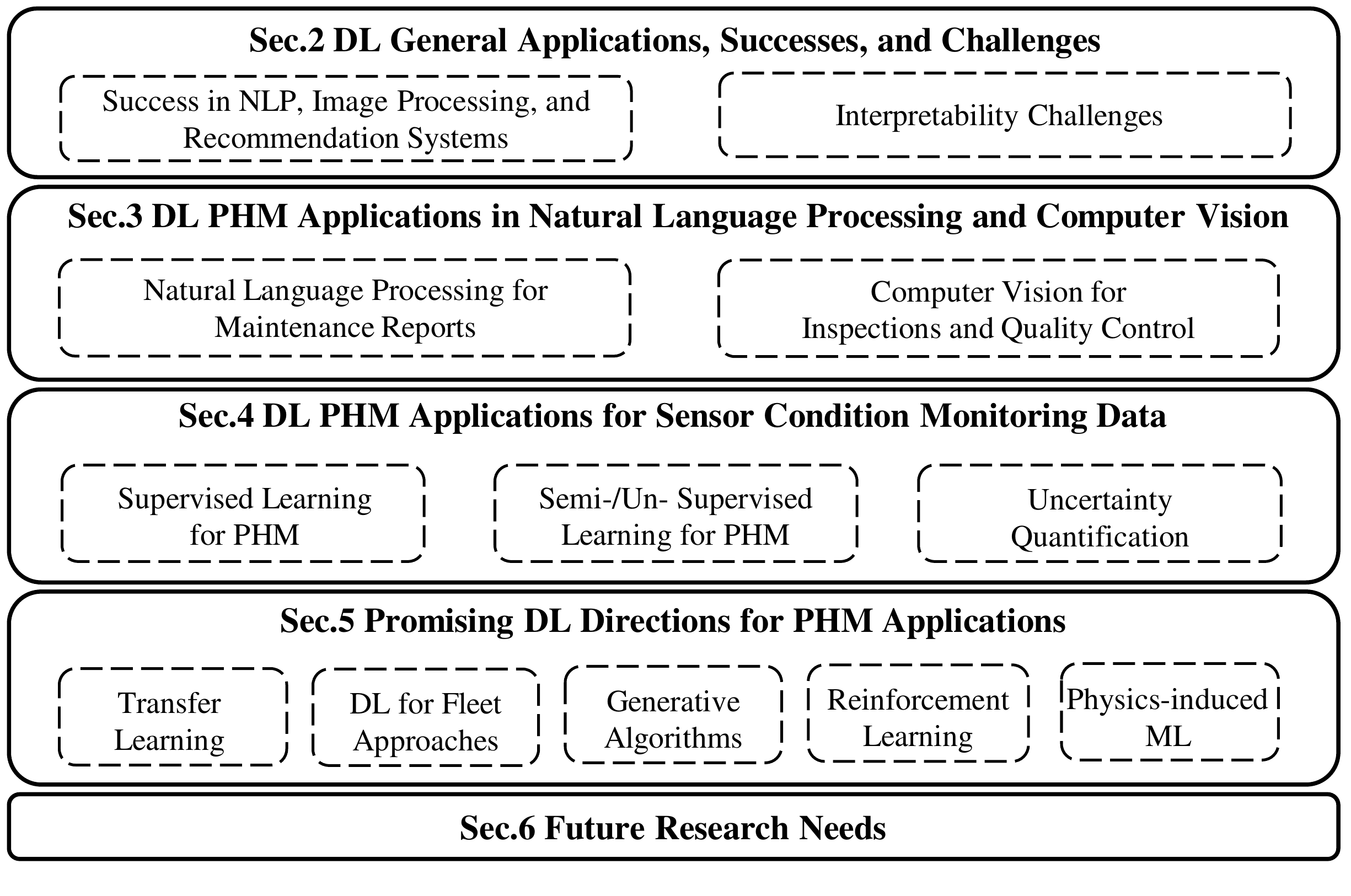}
\label{fig:sections}
\end{figure}

This review paper provides a thorough evaluation of the current developments, drivers, challenges and potential solutions in the field of DL applied to PHM applications. Due to the wide variety of possible topics and potential directions, we focused on the most promising aspects and directions with the potential of solving the current challenges in industry as pointed out in the above section. The content and structure of the paper are shown in Figure \ref{fig:sections}. 

\section{Deep Learning Applications, Successes and Challenges}

\subsection{Introduction to Deep Learning}

Deep learning is an overarching concept that encompasses new variants of a range of established learning models, known as neural networks \cite{DBLP:books/lib/Bishop07}, now more commonly referred to as deep neural networks (DNNs) \cite{Goodfellow-et-al-2016,zhang2019dive}. DNNs are called "deep" because they have multiple layers of computational units, whereas traditional, "shallow" neural networks usually only have a single layer. The extension from a single layer to multiple layers may seem rather obvious and various forms of DNNs were indeed tried already during the early--mid 1990s \cite{Tesauro:1992:PIT:139611.139616}. However, these early attempts at training DNNs did not seem to offer any advantage compared to single-layer neural networks, rather the opposite, and consequently interest in this direction of research diminished. With time, though, new ways of training DNNs were gradually developed \cite{DBLP:conf/nips/BengioLPL06,Lecun98gradient-basedlearning,Hinton2006AFL}, while the performance of computer hardware continued to improve. In 2012, there was a breakthrough, in terms of reaching beyond the neural networks and machine learning communities, when a deep convolutional neural network (CNN) set new top scores in image object recognition \cite{NIPS2012_4824, DBLP:journals/corr/HeZRS15}. This resulted in a surge of interest in DL, benefiting research that in many cases was already well underway, and resulting in rapid advancements of fields of applications for DL.

A common theme of DL is that it is used to tackle complex problems such as computer vision or NLP, for which near-human performance seemed a long way away a little more than a decade ago. These problems are tackled using intricate models with very large numbers of parameters, which are trained on very large datasets with only a limited amount of pre-processing required. Often, the architecture of these models reflects assumed structure in the data. For example, deep CNNs are typically used for image processing (Section \ref{sssec:imgproc}), whereas recurrent neural networks (RNNs) are often used for sequential data, such as NLP data (Section \ref{sssec:nlp}). This structural match between model and data facilitates the identification of relevant features in the data during the learning process and DL models are frequently claimed to circumvent the need for feature engineering \cite{Forman2003, bengio2013representation}. While several factors have contributed to the explosive growth of deep learning, three stand out as being particularly important:
\begin{itemize}
\item Availability of large labeled datasets required to train large DL models with large numbers of parameters \cite{deng2009imagenet}. Some of these datasets have been built over considerable time whereas others have been created through internet-scale crowd-sourcing efforts \cite{fiedler2018imagetagger}, while others yet are products of an increasingly digitized society. Many of these datasets are publicly (though not necessarily freely) available on the internet and several are used as benchmark datasets; some of them will be mentioned (with references) in the following sections.
\item Emergence of fast, affordable hardware \cite{Coates2013DeepLW}, well suited to carry out the large amount of computation required in training DL models. While this hardware was initially developed for the gaming console/PC mass-market, it is nowadays available specifically adapted for the purposes of DL~\cite{haensch2018next}. 
\item The development of several sophisticated, freely available software libraries \cite{Abadi2016TensorFlowAS,DBLP:journals/corr/JiaSDKLGGD14,Seide:2016:CMO:2939672.2945397,AlRfou2016TheanoAP,paszke2017automatic} has made building, training and evaluating DL models relatively straightforward. Although highly complex, several libraries include or are accompanied by wrapping libraries that make the creation of models with established architectures simple and intuitive, requiring only a small amount of coding. 
\end{itemize}
The need for large computational efforts and, especially, large labeled datasets may be too difficult or costly to satisfy for many learning problems. Fortunately, it has been shown that DL models already trained on one task can often be retrained to perform a similar task. Such retraining, commonly known as \emph{transfer learning} \cite{Yosinski2014HowTA}, typically requires much less data and computational resources, making DL applicable to a potentially much wider range of practical tasks. 

\subsection{Deep Learning Successes}

Before looking at the current and future applications of DL within the field of PHM, we first review some of the most significant successes DL has achieved so far in other fields. These serve as inspiration as well as examples for how to move the field forward.

\subsubsection{Speech Recognition, Machine Translation and NLP} 
\label{sssec:nlp}

Many people probably do not realise that they already own one or more devices that implement DL technology. The speech recognition systems on modern smartphones, tablets and computers were possibly the first widely available applications of DL. These first systems used deep belief networks (DBNs) \cite{HintonEtAl_DNN_AcoustMod_2012} in conjunction with hidden Markov models (HMM) \cite{Rabiner89atutorial} to deliver a significant performance improvement on the until then prevailing technology, which had used Gaussian mixture models \cite{DBLP:books/lib/Bishop07} in place of the DBNs. More recent DL based systems have moved on to use variants of RNNs \cite{Amodei2015DeepS2,He2018StreamingES} in so-called end-to-end systems that implement the complete speech recognition process within a single model. Most recent speech recognition systems have been trained on large, proprietary datasets, but there is a number of publicly available datasets such as TIMIT \cite{timit} and Switchboard \cite{switchboard}, published by the Linguistic Data Consortium (LDC)\footnote{\texttt{https://www.ldc.upenn.edu/}}, which have been frequently used to benchmark speech recognition models.

DL has also had a significant impact on processing of natural language in written form. Various types of RNNs have successfully been applied to the task of machine translation \cite{Kalchbrenner2013RecurrentCT,Sutskever2014SequenceTS,Wu2016GooglesNM,Hieber2017SockeyeAT}, achieving performance that is comparable to or exceeding that of carefully handcrafted models. RNNs have also been shown as being capable of capturing aspects such as style of writing and syntactic rules from raw text data ranging from Shakespeare plays to the C source code of the Linux kernel \cite{KarpathyFunRNNs}. While these results are remarkable, a popular and possibly better alternative to raw text data is offered by new representations for language components, such as the word2vec embedding of words \cite{Mikolov2013ICLR}, which encodes words as vectors, such that words that share aspects of their meaning will be mapped to similar vectors. Embeddings at the character and sentence level have also been proposed. Using this kind of representations, DL models have been successfully applied to classic NLP tasks, such as part-of-speech tagging and named entity recognition, as well as sentiment analysis, and question-answering and dialogue systems \cite{Young2017RecentTI}\footnote{As is the case of speech, text data for different tasks are published by LDC and machine translation data can be found from \texttt{http://statmt.org/}}. 

\subsubsection{Image Processing} \label{sssec:imgproc}

In the last decade, deep neural networks, in particular deep CNNs \cite{Lecun98gradient-basedlearning}, have revolutionised the field of image processing. These models were proposed as early as 1989 \cite{LeCun1989HandwrittenDR} for the purpose of image classification and were trained to recognize images of numbers from handwritten ZIP-codes collected by the US post office, the so-called MNIST dataset \cite{LeCun2005TheMD}. However, it was only in 2012 that these models truly caught the attention of wider research communities, after a deep CNN, subsequently named AlexNet, established a new top score in object recognition \cite{NIPS2012_4824}. The model was trained on the then recently released ImageNet \cite{imagenet_cvpr09} dataset, which contains more than a million images that have been labeled using a crowd-sourcing effort. There are also several smaller datasets, such as CIFAR \cite{Krizhevsky2009LearningML} and Caltech-101 \cite{Li2004LearningGV}, that have been  used extensively  in DL research. The AlexNet paper was followed by several papers \cite{Simonyan2014VeryDC,Szegedy2014GoingDW,DBLP:journals/corr/HeZRS15}, exploring increasingly deeper and intricate network topologies, with in some cases more than hundred layers and millions of parameters. Focus has also moved on from simple object detection to object instance segmentation, where multiple objects, including individual instances of the same category, are segmented out at the pixel level \cite{He2017MaskR}. Researchers have also tackled joint NLP-image-processing problems by combining RNNs and CNNs into systems for automatic generation of captions to images \cite{Karpathy2015DeepVA}.

\subsubsection{Recommendation Systems}

Many of the remarkable successes of DL have been in applications where humans have traditionally excelled above computers, such as those described above. However, it has also been used to successfully tackle problems that have arisen out of modern Internet technologies, such as large-scale recommendation systems \cite{Adomavicius2005TowardTN,Ricci2011IntroductionTR}. As the name suggests, these systems seek to provide recommendations on content to present to users, who are often acting in the role of consumers. This enables automatic personalization of the recommended content, product or services to a large number of users. An example are the additional items offered on sites such as Amazon.com once a user has bought, or merely expressed an interest in an initial item; another are the film recommendations users receive on Netflix \cite{Bennett07thenetflix}. Common to all such systems is that they learn user preferences from large, often vast collections of data. Hence, they are well-suited to the application of DL technologies \cite{Zhang:2019:DLB:3309872.3285029}. For example, DNNs have been used to improve the recommendation systems for the YouTube online video content platform \cite{Covington2016DeepNN}, providing suggestions to users on what they might want to watch. Similarly, DNNs have been combined with generalised linear models to improve the recommendation system for apps in the Google Play mobile app store \cite{Cheng2016WideD}. These successes have the potential to be transferred to a PHM context. For example there are some parallels between recommendation systems and decision support systems that, i.e. in the context of suggesting actions for an operator, in order to benefit from both the expertise of artificial intelligence and experienced humans. However, decision support systems are not only used to recommend the optimal actions but also they can also suggest system operational set points, i.e. to prolong the remaining useful life. Although these scenarios seem to have a lot of commonalities, there are also challenges. For example, the actions used in PHM applications are often continuous instead of discrete, making it hard to directly adopting classical recommendation system algorithms to PHM applications. Also, recommendation systems generally rely on an abundance of data which is rarely present in the PHM context. 



\subsection{Interpretability}


The inherent characteristics of deep neural networks, particularly the non-linear computations that are distributed over single nodes and layers, make them on the one hand powerful in terms of computational performance and on the other hand they do not allow for any interpretability. 
Therefore, particularly the users and the domain experts in the specific fields have been concerned about the interpretability, explainability, transparency and understanding of the models and the results. Also, the term “trust” has been recently increasingly raised in the context of deep learning. Whereby “trust” can be related to different topics, including bias, lacks of representativeness of the training datasets and adversarial attacks. Trust can be developed or enhanced by improving the transparency, the interpretability and the understanding of the algorithms \cite{Lipton2016TheMO}. 
The focus of this section is particularly on the term of interpretability which is also used as a synonym to explainability. The goal is to understand and explain what the model predicts. One approach could be to build interpretability into the model, e.g. by fusing physical models and machine learning or learning the underlying physics explicitly, as described in the section~\ref{fig:Physics_informed}. Here, however, we focus on an alternative, post-hoc approach to interpretability \cite{Montavon2018}, that tries to explain given models and their output.
%
This approach seeks to provide interpretability on different levels \cite{Samek2019}:  

\begin{itemize}
  \item Explaining learned representations by analyzing the internal representation of (e.g.) neural networks
\item Explaining individual predictions 
\item Explaining model behavior. 
\end{itemize} 
Furthermore, different approaches have been proposed to explain the behavior of machine learning approaches post-hoc, these include \cite{Samek2019}: 
\begin{itemize}
\item Explaining with Surrogates 
\item Explaining with local perturbations 
\item Propagation-Based Approaches (Leveraging Structure) 
\item Meta-explanations.
 \end{itemize}
Particularly surrogate approaches have been gaining popularity where simple surrogate functions are used to explain the behavior of more complex models. One of such models is the Local Interpretable Model-agnostic Explanations (LIME) \cite{Ribeiro2016}. 
Most of approaches proposed to improve the interpretability are particularly applicable for computer vision tasks where the interpretability and understanding can be gained by different types of visualizations. While some of the approaches can be also transferred to the tasks in prognostics and health management, the explainability may not be sufficient for physical systems.

\section{DL PHM Applications in Natural Language Processing and Computer Vision}
\subsection{Natural Language Processing for maintenance reports}
While sensory data is perhaps what most people would first think of when hearing the term "PHM data", there are other kinds of information that can be as important and are often available in PHM. Failure/Incident reports, maintenance records and event logs can help identify frequency, causes and remedies of failures, but require different processing than (numerical) sensor data. These data may be used as sources for both input data and labels/targets.  However, in most cases, failure records and maintenance reports are entered manually by technicians as free form text, which introduces a certain degree of variability and sometimes even ambiguity. NLP techniques can help to automate this process and get more insights from these data, e.g. to identify the root causes of failures.

Current DL models for NLP have two major components, i.e., representation learning and RNNs. Representation learning in NLP is also called word embedding \cite{Mikolov2013ICLR,Peters:2018}. Learning of word embedding can be done by either adding an embedding layer to a deep neural network or outside the deep neural network by unsupervised learning. Word embedding performs automatic feature engineering that can capture the correlations between consecutive words as well as between words and the wider context they appear in. These representations typically are more compact than the other widely used 1-of-$N$ representations of words. Long short-term memory (LSTM) \cite{Hochreiter1997NC} is one of the most popular RNNs used in NLP. With the memory mechanism, it can capture not only short term but also long term relationships within the text. 
Integration of word embedding and RNNs enables end-to-end learning and makes the adoption of NLP in PHM a viable proposition. With the help of NLP, we can identify keywords in maintenance records or failure reports and predict the type of failure a maintenance record remedied. Recently, NLP techniques have been used in PHM for various tasks.  For example, Su et al. \cite{su2019deep} uses LSTM models to provide fault diagnosis for storage devices based on self-monitoring logs.  Ravi et al. \cite{ravi2019substation} conduct substation transformer failure analysis based on event logs.

\subsection{Computer Vision for Images and Videos from Inspections and Quality Control}

Many PHM applications based on DL on image data have been proposed in recent years, typically with the motivation of automating the processing of these data, e.g. classifying or deriving quantitative descriptors for the content of these images. Gilbert et al \cite{Gibert2015DeepML} built a multi-task deep CNN using image data collected for the purpose of monitoring railway infrastructure. The use of dedicated equipment allowed a substantial amount of data to be collected along railway track, which were manually labeled using purpose-built software. The multi-task CNN was capable of recognizing  parts of railway infrastructure as well as detecting damage such as cracked concrete ties and missing or broken rail fasteners. Liu et at \cite{Liu2018AIfacilitatedCC} used image data collected with drones to train a CNN to detect coating breakdown and corrosion (CBC) in ballast tanks of marine and offshore structures. They employed a transfer learning approach using a VGG19 model \cite{Simonyan2014VeryDC} that was re-trained to classify different kinds of CBC. Similar image monitoring schemes have also been proposed in the field of aviation. Two different ImageNet-trained CNNs are compared in \cite{Malekzadeh2017AircraftFD} as feature detectors, the output of which is combined using an SVM to detect defects in aircraft fuselage. Svens\'{e}n et al \cite{Svensn2018DeepNN} use VGG16-based CNN models \cite{Simonyan2014VeryDC} to process images of jet engines and jet engine parts. In a first step, a CNN is used to separate conventional camera photographs from images collected during borescope inspections of engines. The latter are subsequently analysed for part detection using another CNN. For the first task, a substantial proportion of the available images could be reliably labeled using auxiliary labels or reliable heuristics whereas for the second task, images were manually labeled. Jet engine parts, in particular, turbine blades are also the focus of Bian et al \cite{Bian2016MultiscaleFC}, who used a fully convolutional neural network  to process images of turbine blades that were collected in a dedicated imaging rig, following removal from the engine. This more elaborate data collection process enabled accurate segmentation of regions where the protective coating of the blade had been (partially) lost, which can be used as an indicator of the condition of the blade.

\section{DL PHM applications for sensor condition monitoring data}

\subsection{Supervised Learning Algorithms for PHM applications}
Supervised learning is the machine learning task of learning a function that maps an input to an output based on example input-output pairs~\cite{friedman2001elements,stuart2003artificial,Goodfellow-et-al-2016}. Classic supervised tasks in other fields include image classification~\cite{krizhevsky2012imagenet,he2016identity}, and text categorization~\cite{cavnar1994n,sebastiani2002machine,johnson2017deep}. To tackle these tasks, researchers in other fields have developed various kinds of deep learning structures. Some of them are particularly useful for PHM applications. 

\paragraph{Recurrent structures}
Recurrent neural networks~\cite{hopfield1982neural, rumelhart1988learning, graves2013speech, graves2013generating}, especially  LSTM~\cite{gers1999learning, graves2005framewise} networks have been widely used for PHM applications~\cite{yuan2016fault, malhotra2016multi, zhao2016machine, zhao2017learning, wu2018remaining}. After the invention of LSTM, this line of work has been heavily developed in other fields, especially natural language processing. The method was further improved by bidirectional LSTM~\cite{graves2005framewise}, and Gated Recurrent Unit (GRU)~\cite{heck2017simplified, dey2017gate}. In recent years, the \emph{attention mechanism}~\cite{zhou2016attention} has improved the performance of LSTM networks on time series data significantly, especially on language tasks. Networks trained using the attention mechanism~\cite{vaswani2017attention, devlin2018bert, radford2019language} currently hold the state-of-the-art performance on a series of language tasks. This recent progress, however, is not yet fully exploited for PHM applications.

\begin{figure}[hbt!]
\caption{Examples of 2D-encodings of time series with GAF, MTF and RP}
\centering
\includegraphics[width=0.6\textwidth]{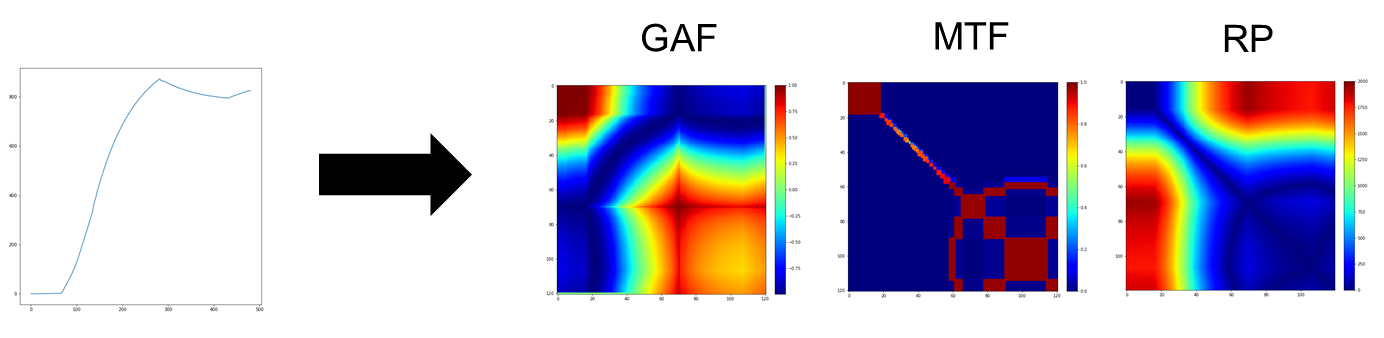}
\label{fig:Image_encodings}
\end{figure}

\paragraph{Time series to image encoding}
Motivated by the success of CNNs on image representation learning~\cite{deng2009imagenet, he2016identity}, there has been a rising trend on understanding time series by translating them into images. By doing so, existing knowledge on image understanding and image representation learning can be directly exploited for PHM applications. An intuitive approach~\cite{krummenacher2017wheel} is to simply use the natural plot of 1-D time series data, signal vs time, as two-dimensional image. Alternatively, as shown in Figure~\ref{fig:Image_encodings}, Gramian Angular Fields (GAF), Markov Transition Fields (MTF) and Recurrence Plots (RP) have been introduced in~\cite{wang2015encoding, Debayle2018} as encoding approaches to translate signals to images. These encoding methods were adopted by~\cite{gecgel2019gearbox} for vibration data failure detection. Furthermore, time-frequency analysis also results in two-dimensional signal representations. The evaluation of the benefits of the different time series to image encodings and the comparison of their performance for different types of time series data and different fault types is a currently an open research question~\cite{Rodriguez2020}. 

\paragraph{1-D CNN}
An alternative approach is to accept the fact that sensor data are often 1-D data by nature instead of 2-D images. In order to still benefit from recent progress in convolutional operations, 1-D CNN kernels, instead of 2-D kernels, can be used for time series classification, anomaly detection in time series or RUL prediction. The same idea has been widely adopted for motor fault diagnosis~\cite{ince2016real}, and broader PHM applications~\cite{babu2016deep, jing2017convolutional, li2018cross, wang2019domain}. 

\paragraph{Combinations of LSTM and CNN}
Both RNNs and CNNs have shown their applicability in understanding time series for PHM applications. It is thus natural to leverage the power of both RNNs and CNNs. Existing works~\cite{canizo2019multi, zheng2019fault} in PHM mainly adopt a sequential approach, which first extracts local features using CNNs and then feeds them to LSTMs for temporal understanding. An alternative solution~\cite{xingjian2015convolutional, stollenga2015parallel} is to embed convolution within the LSTM cell, such that convolutional structures are in both the input-to-state and state-to-state transitions. This approach has been widely adopted in video understanding~\cite{lotter2016deep, Byeon_2018_ECCV} and volumetric medical data analysis~\cite{chen2016combining,litjens2017survey}. Recently,~\cite{zhang2019deep} adopted this idea and proposed to use ConvLSTM for anomaly detection and fault diagnosis on multivariate time series data. 

\subsection{Unsupervised and semi-supervised learning  for PHM applications}
Supervised learning requires labels or supervision for every training sample. However, this can be too strong a requirement, as labels can be expensive or even infeasible to collect. In reality, we would also like to learn partly or fully from unlabeled data. This leads us to the field of unsupervised and semi-supervised learning.
In PHM applications, a label can constitute either a health indicator (for fault detection tasks), a specific fault type (for fault classification tasks), or the RUL at each time step of measurements (for prognostics tasks).
Particularly in safety critical systems or in systems with high availability requirements, faults are rare and components are replaced or refurbished preventively before reaching their end of life. Thereby, the lifetime of the components is truncated and the true lifetime remains unknown. Furthermore, assessment of the true system health is often too expensive or may be impracticable. This results in a lack of labels for many PHM applications.

Therefore, labels cannot be used to learn the relevant patterns. Other approaches need to be developed to tackle these PHM challenges. These are mostly unsupervised and semi-supervised learning approaches.

One of the most popular unsupervised learning approaches applied in PHM is \textbf{signal reconstruction}. The general idea behind the signal reconstruction is to define a model that is able to learn the normal system behaviour and to distinguish it afterwards from system states that are dissimilar to those observed under normal operating conditions. This technique is also known as novelty or anomaly detection. Generally, the implementation of signal reconstruction can be performed either with physics-based but also increasingly with data-driven approaches. In this context, \emph{autoencoders} \cite{Hinton1993AutoencodersMD} are typically applied. Particularly for the data-driven signal reconstruction approaches, it is important to identify representative system conditions and to use condition monitoring data from these conditions to train the algorithms and learn the underlying functional relationships. The residuals are then used to detect the abnormal system conditions. Deep autoencoders \cite{DBLP:conf/nips/BengioLPL06} can potentially capture more complex relationships between the condition monitoring signals and may hence be able to detect more subtle deviations from the representative system conditions. Since the autoencoders are an unsupervised learning architecture, they do not require large amounts of data to be labelled.

Clustering, adaptive dictionary learning \cite{Lu2019} or causal maps combined with multivariate statistics \cite{Chiang2015a} are some examples of other unsupervised learning approaches applied for fault detection but also for fault isolation. Traditionally, clustering was applied directly to the preprocessed condition monitoring data or the handcrafted features \cite{Detroja2006, Hu2012, Li2014, Du2014, Costa2015, Zhu2017}. Clustering approaches, particularly adaptive density clustering algorithms, can distinguish well the different fault types if the feature space (i.e either the raw CM data or manually extracted features) is separable in fault types. However, distinguishing between fault types becomes increasingly difficult when there is a large number of CM signals, i.e. when the dimensionality of the input space increases and when these signals are additionally noisy and correlated. In such cases, it is often difficult to interpret distances in the feature space and, therefore, to define a valid proximity or similarity metric. As a result, some fault types can become inseparable, leading to clusters with mixed fault types. Therefore, recently clustering was applied on the more informative latent feature representation \cite{Yoon2017}. In this case, again autoencoders are first applied to compress the high-dimensional condition monitoring data into lower dimensional representation of the input space and then clustering is performed on this informative latent space.

When working with large data sets where only small subsets of the samples are labeled, \textbf{semi-supervised learning} approaches have been applied \cite{oliver2018realistic, zhu2003semi, Chapelle2006, Zhu2009}. Semi-supervised learning leverages the available unlabeled data to improve the performance of the supervised learning task. Different concepts have been proposed for semi-supervised learning tasks. These include (deep) generative models \cite{KingmaSemi}, graph-based methods \cite{Zhao2015} and transductive methods \cite{bruzzone2006novel}, to name a few. A further possibility to distinguish the different semi-supervised learning approaches is to differentiate between those based on consistency regularization~\cite{lee2013pseudo, tarvainen2017mean, miyato2018virtual}, entropy minimization and the traditional regularization \cite{zhu2003semi, Chapelle2006}. We refer interested readers to the review paper~\cite{oliver2018realistic}.


Recently, motivated by its successful use in supervised learning, the MixUp \cite{zhang2017mixup} method, mixing labeled and unlabeled data, has been found to be another promising approach to semi-supervised learning~\cite{berthelot2019mixmatch, verma2019interpolation, wang2019semi}. 

Some of the semi-supervised approaches have also been applied to PHM applications \cite{Shi2018, ListouEllefsen2019}, including self-training \cite{Yoon2017}, graph-based methods \cite{Zhao2015} and co-training methods \cite{Hu2011}. Most of the contributions on semi-supervised learning for PHM applications have been focusing on the prediction of the RUL \cite{Yoon2017, ListouEllefsen2019}.

In situations where unlabeled data is abundant but acquiring corresponding labels is difficult, expensive or time-consuming, the concept of active learning has been flourishing. Active learning \cite{Settles2009} enables machine learning algorithms to achieve a higher accuracy with less labelled data, provided the algorithms are allowed to choose the training samples to be labelled out of the large pool of unlabelled data. Chosen data are labelled by "querying an oracle", which might be a human expert or another method that affords labelling only a subset of the available data \cite{sener2017active}. 

Fault detection can also be defined as a \textbf{one-class classification}  problem \cite{Michau2017}. However, differently to previously proposed one-class formulations \cite{Scholkopf2000}, the problem is defined in \cite{Michau2017} as a regression problem where the healthy system conditions during training are mapped to a constant target value $\mathbf{T}$.  If presented with operating conditions that are dissimilar to those observed during training, the output of the trained network will deviate from the target value. The problem formulation can be also considered as an unbounded similarity score to the training dataset with known healthy system conditions.

\subsection{Uncertainty quantification and decision support}

In PHM, especially prognostics, it is of paramount importance to take into account uncertainty inherent in models and data. Predicting RUL as a single number, for instance, is illusory and can be deceptive. Decisions made on the basis of a deterministic model for variables that contain randomness are bound to be flawed.

Sources of uncertainty are several.  First, the uncertainty inherent in the models, or epistemic uncertainty: physics-based  degradation models are approximations of reality and usually contain unknown parameters. Second, any data acquired through sensors are affected by measurement errors: this is measurement uncertainty. And, last but not least, the future operating profile and loading of equipment being monitored is not known with certainty, yet it is key in predicting the evolution of degradations and how soon they can lead to failures.

As a result, the RUL, in order to be meaningful, must at the very least be accompanied by confidence intervals and, which is even better, by a description through probability distributions if at all possible, or by fuzzy representations.
The case is made in \cite{Sankararaman2015} that a sharp distinction must be operated between test-based health management and condition-based health management: the former, being based on a population, can resort to more classical (frequentist) statistical techniques, while the latter, focused on one single asset, can only rely on Bayesian techniques. This is  because, in the case of one single asset, variability among specimens does not occur.

The challenge is to propagate over time the uncertainties on both model and asset initial state. The distribution over future states and RUL are the outcomes of uncertainty propagation and should not be assumed to have a certain form (for instance, assumed to be Gaussian  without any justification).
In most cases, the only practical method to estimate the uncertainty in the RUL is through Monte Carlo simulation and uncertainty propagation because, most of the time, the RUL is a nonlinear function of the initial state, and therefore, will not follow a Gaussian distribution even if the initial state does.
Prognostics is only useful if it can aid decision making.  According to what precedes, one has to deal with decision making under uncertainty.

Beyond uncertainty propagation, the need therefore arises for uncertainty management. For instance, if the variance of the RUL  is very large, how can one deal with the uncertainty in input conditions to reduce the uncertainty on the RUL. 

A survey of uncertainty propagation techniques in DL  has been published in  \cite{Abdelaziz2015}. Applications involve image processing  and  natural language processing.

Not surprisingly, given the nonlinear nature of DNNs, Monte Carlo simulation is the preferred method for propagating uncertainty; but, in view of the computational cost of simulations, approximations are being sought, such as responses surface techniques \cite{Weiqi2019} which rely on the property that DNN outputs are influenced by just a few inputs and the sensitivity to various inputs can be assessed by evaluating gradients.

An interesting approach, introduced by MIT in this context  \cite{Titensky2018}, is the use of the extended Kalman filter (EKF): if  the layer in the deep network is used as the time step and the value as the state, it is possible to propagate uncertainty in the initial input but also process noise (the latter results from errors in biases and weights of the pre-trained network). 
In spite of what has been stated above in terms of frequentist versus Bayesian approaches, useful insights can be derived  from consideration of the mean residual life (MRL) as the expectation of RUL  (the expectation can be seen as Bayesian but possibly also as frequentist if a large number of  similar scenarios are considered or if many identical assets in a fleet are considered). Recent work \cite{Huynh2017} has made progress toward  including a probabilistic description of RUL in maintenance optimization approaches.
Also, in some categories of problems \cite{Dersin2018}, it has been shown that the  variability of RUL, measured by  a coefficient of variation, decreases with the rate of ageing (i.e. the rate at which the MRL decreases as a function of time), and useful methods of traditional reliability engineering can be brought to bear on uncertainty quantification of the RUL \cite{Dersin2019}.
A typical example of decision support problem that relies on PHM results is that of dynamic maintenance planning: deciding when preventive maintenance operations should be performed on assets of a fleet (such as trains or aircraft) on the basis of RUL predictions communicated by the various assets and on logistic and operational constraints (such as spare part management policies and schedule adherence constraints). See \cite{Herr2017} for a railway application that involves multi-agent modelling.

\section{Promising DL Directions for PHM Applications}

\subsection{Transfer Learning} 

 Most DL methods assume that training and test data are drawn from the same distribution. However, the difference in operating conditions often leads to a considerable distribution discrepancy. Consequently, applying a model trained on one machine to another similar machine that is operated differently may result in a performance drop. Therefore, to apply a DL model for a fleet of machines, collecting, labeling, and re-training models for each machine, is a series of necessary but tedious jobs. Unsupervised domain adaptation techniques, as a subtopic in transfer learning, provide a promising solution to this problem by transferring knowledge from one well-understood machine to another in the fleet, without the need of labels from target machines. 
 
In addition, we are facing more challenges when adopting transfer learning or domain adaptation methods from a general setup to the industrial settings. For example, in the ideal general setup for domain adaptation,  the source and target tasks have the same input space and same output space. This is particularly not practical in industrial setups, where the set of input sensor signals is likely to be different across different machines, and the set of output labels(types of faults, RUL range) may also be different between different machines. Limited number of existing research works~\cite{8949730, 8917808} focus on the case where the output label space is not identical in source and target. 

Before the wide adaptation of deep learning approaches, domain adaptation was mainly tackled by learning invariant features across domains~\cite{saenko2010adapting,  pan2010domain, duan2012domain, zhang2013domain}, or assigning weights to source samples based on their relevance to the target~\cite{gong2012geodesic}. Before the rise of adversarial training, distribution alignment approaches by statistics matching~\cite{long2015learning, yang2016revisiting, carlucci2017autodial} focused on using statistical moments to mitigate the domain difference. In recent years, adversarial approaches~\cite{ganin2014unsupervised, tzeng2017adversarial, ganin2016domain}, where a separate discriminator is used to align the distributions, are yielding superior performance on many different tasks. 

Domain adaptation methods have recently also been introduced to PHM applications, especially on fault diagnosis problems. AdaBN~\cite{zhang2017new}, adversarial training~\cite{zhang2018adversarial, wang2019domain}, and  MMD-minimization~\cite{li2018cross, li2019multi} were used to align the full source and target distributions for rotating machines. 
In addition to these domain adaptation approaches, there is a recent trend of direct transfer learning without explicit distribution alignment in PHM applications. These works~\cite{shao2018highly,cao2018preprocessing} convert raw signals into graphical images as input, and then fine tune ImageNet pre-trained image classification models for their own fault diagnosis tasks. Despite its  simplicity, this approach leads to promising results on fault diagnosis for gearbox, and induction motors. 

In summary, many domain adaptation and transfer learning methods have been proposed in recent years. However, the setups used by these methods need to be refined in order to meet the need of PHM applications. How do we deal with the case where input and output space of the source and target domain are not identical? How can we efficiently encode the input data to make them more transferable? How do we deal with the quantity imbalance between  healthy and fault data? These are the questions which require more research. 

\subsection{DL for fleet approaches}
Failures occurring in critical systems, such as power or railway systems, are rare and the fault patterns are often unique to the specific system configurations and specific operating conditions. Therefore, they cannot be directly transferred to a different system of the same fleet operated under different conditions. There are different perspectives on the fleets of complex systems. The most commonly used perspective is the one where a fleet is defined as a set of homogeneous systems with corresponding characteristics and features, operated under different conditions, not necessarily by a single operator \cite{Leone2017}.  A good example is a fleet of gas turbines or a fleet of cars produced by one manufacturer with different system configurations, operated under different conditions in different parts of the world. The biggest challenge in the fleet PHM is the high variability of the system configurations and the dissimilarity of the operating conditions. Due to the limited number of occurring faults, the fault patterns from dissimilar operating conditions within the fleet need to be used to learn and extract relevant informative patterns across the entire fleet. Several approaches have been proposed for fleet PHM \cite{michau2019}. For RUL predictions, the existing approaches rely on the assumption that systems that are parts of a fleet all have similar operating conditions and that selected parts of degradation trajectories can be directly transferred from one system of the fleet to another. However, the life of systems within a fleet depends on many different factors, including operating conditions, material properties and variability in manufacturing. Therefore, this assumption is not valid for the majority of complex industrial assets.

This poses a significant challenge on transferring a model developed on one unit to other units of the fleet. Most of the research, has been focusing so far on this perspective, which is already challenging in terms of variability of configurations and operating conditions \cite{Michau2018b}. For the monitoring and diagnosis of such fleets, different approaches have been proposed (organised by increasing complexity) \cite{michau2019}:
\begin{enumerate}
  \item Identifying some relevant operating or design parameters of the units (e.g. average operating regimes) in order to find sub-fleets, possibly with clustering, defined by similar characteristics based on the selected parameters; subsequently using the data of each of the sub-fleets to train the algorithms \cite{Zio2010}.
  \item Using the entire time series of condition monitoring signals to perform time series cluster analysis of units \cite{liu2018cyber}.
  \item Developing models for the functional behaviour of the units and identifying similar units following this learned functional behaviour \cite{Michau2018b}. Contrary to the time series clustering, the approaches do not depend on the length of the observation time period.
  \item Performing domain alignment in the feature space of the different units to compensate for the distribution shift between different units of a fleet \cite{michau2019}.
\end{enumerate}

While the methods used at each complexity level  of the fleet approaches in the list above overcome some of the limitations of the methods used at the previous level, the complexity of the proposed solutions increases. The main limitations of each of these approaches are:
\begin{enumerate}
  \item Aggregated parameters used for comparison may not cover all the relevant conditions or the aggregated parameters may not be representative of the unit specificities.
  \item Comparing the distances between time series is affected by the curse of dimensionality. Time series cluster analysis becomes even more challenging when operating conditions evolve over time.
  \item Even though approaches that learn the functional behaviour of units are more robust to variations in the behaviour of the system, one of the underlying requirements is that the units experience a sufficient similarity in their operating regimes. If the units are operated in a dissimilar way, large fleets may be required to find units with a sufficient similarity (since the similarity is defined at the unit level).
  \item Since the alignment is performed in an unsupervised way and the performance depends on the assumption that the future operating conditions of the unit of interest will be representative to the aligned operating conditions, no guarantees can be made that the system of interest will be behaving in a similar way in the future. However, this limitation is in fact true for all the fleet PHM approaches since the past experience of other fleet units is transferred to the unit of interest. 
\end{enumerate}

Therefore, the challenges for designing a PHM system for fleets from the manufacturer’s perspective are extensive \cite{michau2019} due to the variability of the system configurations and the operating conditions. However, designing a PHM system for a fleet from the  perspective of an operator is even more challenging. In this case, a fleet is defined as a set of systems fulfilling the same function not necessarily originating from the same manufacturer and typically not monitored by the same set of sensors. A good example of such a fleet is a fleet of turbines in a hydro-power plant that may have been taken into operation at different points in time and that were produced by different manufacturers. The units of a fleet share in this case a similar behaviour, fulfil a similar functionality but may be monitored by a different set of sensors, at different locations, different sampling frequencies, etc. This line of research is particularly interesting for the industry. However, this problem has not yet been addressed in data-driven PHM research. Solving this problem will provide a major leap forward for many industrial applications.

\subsection{Deep Generative Algorithms}

Within the last years, Deep Generative Models have emerged as a novel way to learn a probabilistic distribution without any assumption of the induced family distribution. 
Indeed, unlike Gaussian mixtures that assume that the training data are sampled from a mixture of Gaussian distributions whose number of modes and covariance matrices are known beforehand, deep generative models use a neural network, usually denoted as a decoder or a generator, that maps a sample from a noise distribution to a sample from the ground-truth distribution. The deep generative models vary in their implementation mainly from the induced metric distribution used in their optimization scheme.

The most famous example of deep generative models are Generative Adversarial Networks (GANs) (from \cite{goodfellow2014generative}). The training of the Generator $G$ is based on an adversarial loss. The training of the generator is combined with the learning of a second network, the Discriminator $D$ whose goal is to distinguish correctly between real data and generated data. The training will alternate the optimization for both the Discriminator and the Generator until the quality of the generated samples is sufficient.
Despite their success, GANs have suffered from several limitations, mainly due to their training scheme: 
the loss used to train the generator depends on the result of classification of the generated data (does each generated data point fool the discriminator or not?). Thus, once the algorithm discovered how to generate examples that fool the discriminator, the loss leads the training process to generate similar examples and not discover new examples. This leads to mode collapse (generated data only represents a few number of modes from the original data). 

To address this problem, recent research focused on using a metric distribution called the 1-Wasserstein distance to assess the global similarity between generated and original data.
The Wasserstein distance is  based on the theory of optimal transport to compare data distributions with wide applications in image processing, computer vision, and machine learning. Wasserstein Generative Adversarial Networks (W-GAN, see in \cite{arjovsky2017wasserstein}) were first introduced as a solution to the mode collapse problem. Indeed, since the Wasserstein distance is continuous and is a global function, it forces the network not to focus on a subset of the distribution.
Wasserstein GAN learns a generator on a noise distribution (generally Gaussian) so that the output distribution matches the ground-truth distribution. 
In its primal form, the Wasserstein distance requires to measure the expectation of the distances between two continuous distributions which may not be tractable in high dimension. Instead, the formulation of W-GAN relies on the dual expression of the 1-Wasserstein distance, which allows better optimization properties.

Several works apply generative deep modeling, either variational autoencoders (VAE)s or GANs for anomaly detection \cite{an2015variational, schlegl2017unsupervised, deecke2018image}. The principle is as follows: they learn the induced distribution and then assert whether a sample was part of this distribution, by mapping it to the closest sample in the generated distribution. They tackled this mapping, either by adapting an autoencoder learnt during the training like VAEs or DCAEs \cite{an2015variational, masci2011stacked, donahue2016adversarial}, or by doing optimization scheme in the latent dimension directly \cite{schlegl2017unsupervised},\cite{deecke2018image}.

\cite{schlegl2017unsupervised}, \cite{deecke2018image}, whose methods are respectively denoted as AnoGAN and AD-GAN, optimize the latent dimension with gradient descent to approximate at best a given input. If they do succeed in reconstructing the data up to a certain threshold, then the data is recognized as normal. Otherwise, it is detected as anomalous.

When it comes to generating time series data, previous works proposed to use Recurrent Neural Networks for both the generator $\textbf{G}$ and the discriminator $\textbf{D}$ to take into account the sequential nature of the input data. Eventually they generated fake time series with sequences from a random noise space.
Note that those works still requires the duration of the input signal to be fixed.

None of the research studies in \cite{schlegl2017unsupervised}, \cite{deecke2018image} yield an efficient way of estimating the latent value that generated a given input data. Moreover, training GANs is a challenging optimization task, even more, when considering non-image data, so that adding the encoder during the training stage as in \cite{donahue2016adversarial} may harness the generated samples quality.

Recent works by \cite{schlegl2019f} and \cite{ducoffe2019} combine W-GAN and an encoder to perform anomaly detection on medical images and time series respectively.

\subsection{Deep Reinforcement Learning} 
Due to the dynamic and complex nature of corrective and predictive maintenance tasks, it is challenging to react to maintenance requests effectively within a short time. In operations research, scheduling and planning problems are mostly solved by mathematical modelling
if the (simplified) problem can be properly formulated, or by heuristic search if a locally optimal
solution is acceptable. However, neither of these approaches takes into account previous  solutions to similar problem instances. That is, given any problem instance, solutions need to be searched and found from scratch. Deep reinforcement learning \cite{Mnih2015HumanlevelCT} provides potential to adjust to new problems quickly by utilising experience and knowledge gained from solving old problems.

\begin{figure}[hbt!]
\caption{Agent-environment interaction, based on \cite{Sutton1998}}
\centering
\includegraphics[width=0.45\textwidth]{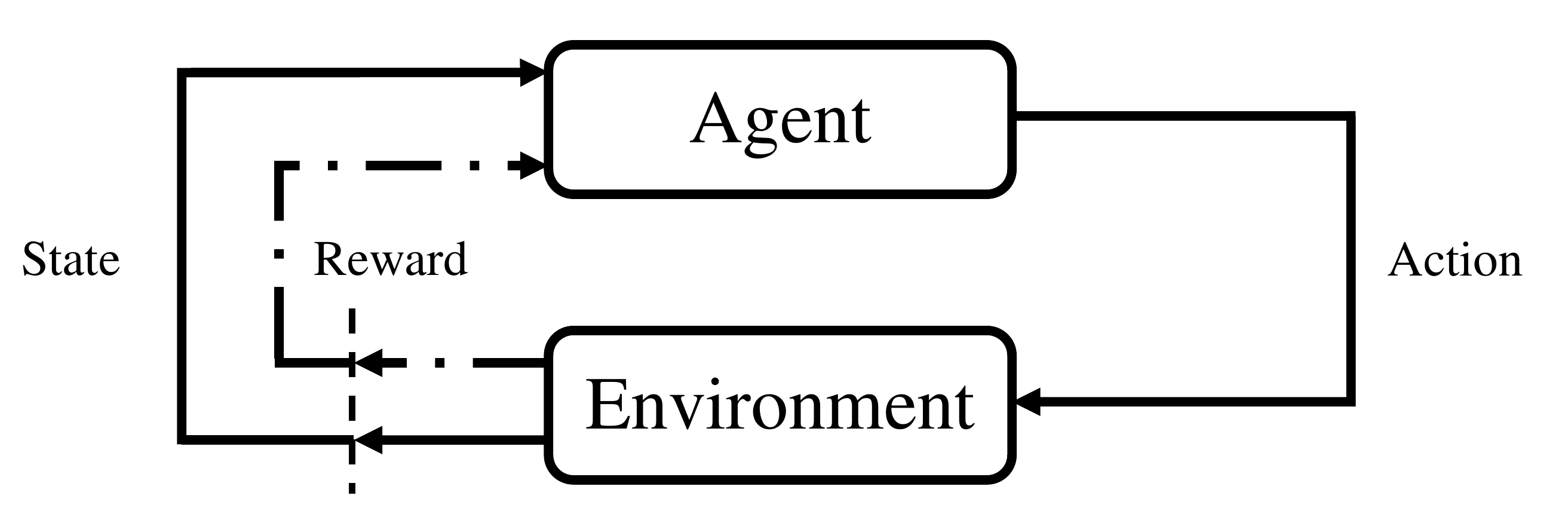}
\label{RLagent}
\end{figure}

Reinforcement learning (RL) systems \cite{Sutton1998} learn a mapping from situations to actions by trial-and-error interactions with a dynamic environment in order to maximize an expected future reward. After performing an action in a given state the RL agent will receive some reward from the environment in the form of a scalar value and it learns to perform actions that will maximize the sum of the rewards received when starting from some initial state and proceeding to a terminal state. By choosing actions, the environment moves from state to state. This interaction is visualized in Figure \ref{RLagent}. Many RL systems have been proposed including Q-Learning (a model-free reinforcement learning algorithm), State–Action–Reward–State–Action (SARSA), and actor-critic learning \cite{Sutton1998}. 

However, for problems with a very large state space, regular RL falls short in learning strategies since the mapping between state-action pairs and rewards is stored in a lookup table. Recent developments in the Deep Reinforcement Learning (DRL), where DL is used as a function approximator for RL, have shown that strategies can be derived for difficult and complex problems. For instance, deep Q-network (DQN), the DRL model that mastered the game of Go  \cite{Silver2016MasteringTG} uses CNNs as a function approximator for the optimal action-value function to replace the Q-table in Q-learning.

DRL  can be for example used in PHM to decide whether to take a maintenance action as shown in \cite{KnowlesBW10}. The age and condition can be used to represent the state of a component and the possible actions will be "maintenance" or "no maintenance". A profit will be returned as a reward if the component does not fail when "no maintenance" is chosen as action. If the component does fail then a repair cost will be deducted from the profit. If "maintenance" is chosen as the action, a maintenance cost is deducted from the profit but there will be no repair cost since the component will not fail. Typically, the maintenance cost is considerably lower than the failure cost. Thus at each time step, the DRL model must decide based on the status of the component between a moderate reward by performing maintenance or risking no maintenance which could incur either a high reward in the event of no failure or a low reward if the component fails.

Another application of DRL in PHM could be using DRL to plan the order of tasks to be handled on assets as given in \cite{Peer2018} to derive a feasible schedule so that all assigned tasks can be finished with the right order in time. 

The major challenge of applying DRL in PHM lies in reward shaping \cite{Marek2017}. Reward shaping is to design a reward function that encourages the DRL model to learn certain behaviors. It is usually hand-crafted and needs to be carefully designed by experienced human experts.  However, it is not always easy to shape the rewards properly and improperly shaped rewards can bias learning and can lead to behaviors that do not match expectations. Moreover, reinforcement learning doesn't learn on fixed datasets and ground truth targets, and therefore could be unstable in its performance.

\subsection{Physics-induced Machine Learning}
A promising approach of inducing interpretability in the machine learning models, particularly for applications outside the image processing domain, where visualizations may not be readily derived, is physics-induced machine learning. Prior knowledge is integrated in the models, delivering improvements in terms of performance as well as interpretability. Several studies have shown that integrating prior knowledge in the learning process of neural networks does not only help to significantly reduce the amount of required data samples but also to improve the performance of the learning algorithms \cite{Karpatne2017,Tipireddy2019}. 

\begin{figure}[hbt!]
\caption{Different ways of integrating prior knowledge in machine learning, derived from \cite{Rueden2019}}
\centering
\includegraphics[width=0.6\textwidth]{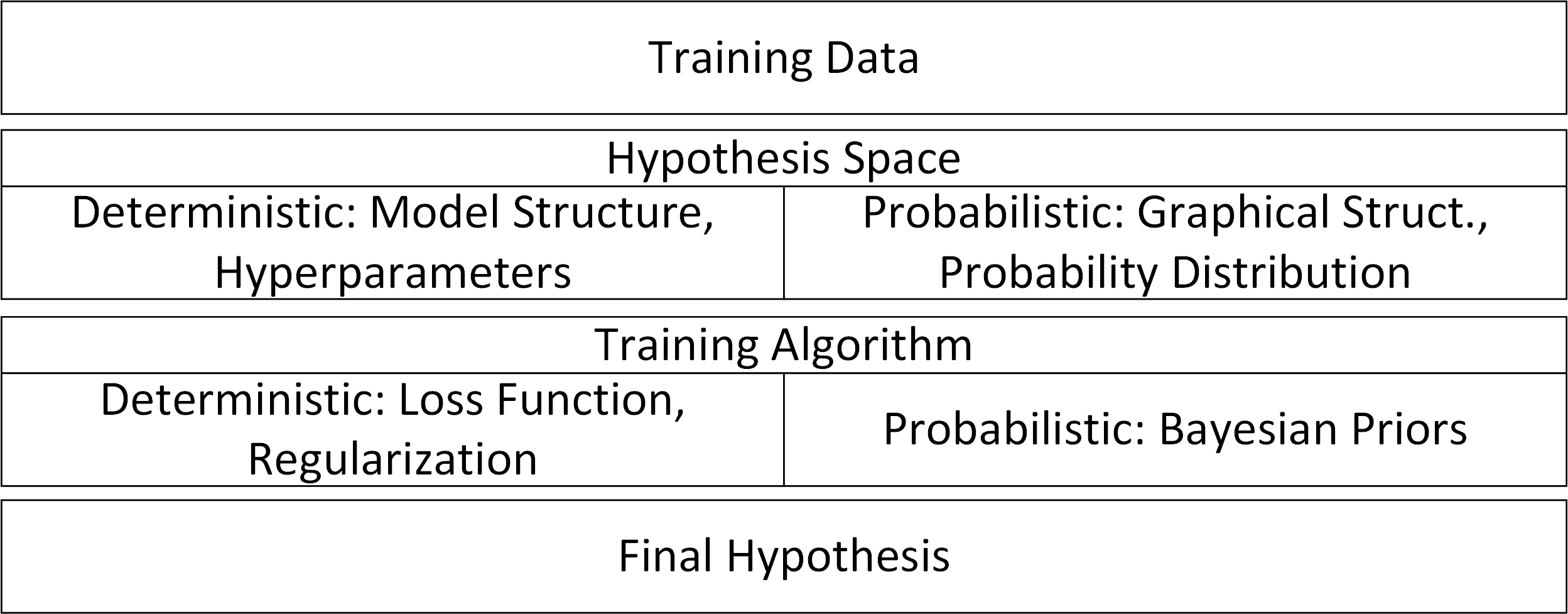}
\label{fig:Physics_informed}
\end{figure}

Different approaches have been proposed for integrating domain knowledge in the learning process, including physics-informed neural networks \cite{Raissi2019, Raissi2018}, physics-guided neural networks \cite{Karpatne2017}, semantic-based regularization \cite{Diligenti2017} and integration of logic rules \cite{Hu16_harnessing}.
As illustrated in Figure \ref{fig:Physics_informed}, integration of the domain knowledge can be performed at different levels of the learning process, such as in the training data, the hypothesis space, the training algorithm and the final hypothesis. In addition, the way how the knowledge is represented and transformed differs quite significantly between the different approaches. It includes simulation, statistical relations, symmetries and constraints  \cite{Rueden2019}. In the field of fault diagnosis and prognosis, the direction of hybrid modelling and physics-informed approaches has only recently been explored with different directions \cite{Nascimento2019, arias19_hybrid, chao2020fusing}. The underlying physical models are integrated to enhance the input space, providing additional information from the physical models to the learning algorithms. Integrating physical constraints and physical models in the AI algorithms will change the model interpretability and also the extrapolation abilities of the algorithms and is subject to further research.

\section{Future Research Needs}

Deep learning provides several promising directions for PHM applications. In terms of methodological advancements, the research directions listed in the previous section: transfer learning, fleet approaches, generative models, reinforcement learning, and physics-induced machine learning hold great potential, but also highlight the need for more research. 

The main focus of domain adaptation research has been on homogeneous cases when source and target input space contain the same features. However, in real applications, complex systems will not only experience a domain shift but will also have different features. In the context of complex industrial systems, heterogeneous domains comprise the case of similar systems produced by different manufacturers and equipped with different sets of condition monitoring sensors (not only sensor locations differ but also sensor types and number of sensors). The research on heterogeneous unsupervised domain adaptation, particularly applied to complex physical systems has been very limited but has a large potential to be impactful, particularly for industrial applications. Another perspective to address the challenges is to make use of simulation environment and adapt from simulation to real-life applications. This direction is particularly interesting as the data will more likely be sufficient in the source domain.

Generative models, have been mainly applied to computer vision tasks. The evaluation of the plausibility of the generated samples in terms of physical system coherence is an open research question. Also, controlling the generation of the relevant data samples from the perspective of the physical processes is similarly an open research question. 

Reinforcement learning has been particularly thriving for applications with extensive and detailed simulation environment. However, this condition cannot be fulfilled for many complex real applications, particularly in the context of PHM. Additionally, the levels of complexity of the problems to which RL has been typically applied in the context of PHM have been still comparably limited. The extension to more complex problems requires additional research.

Combination of DL with expert knowledge is a potentially fertile research area as models can be enriched dynamically with acquired data, thus leading to effective digital twins that can provide  support to  maintenance decision making. While several directions are currently pursued in physics-induced machine learning, there is neither a consensus nor a consolidation on the different directions and how they could be transferred to industrial applications. Further research is required to develop and consolidate these approaches,  which may also result in improvements in the interpretability of the developed models and methods. 

Representative datasets have been key drivers of research and innovation in many leading fields in DL, including computer vision and natural language processing. DL methods require large and representative data sets. However, in the context of PHM, the lack of representative datasets has been obstructing a broad usage and adaptation of DL approaches in industrial applications. There are several solutions to this challenge: data augmentation, data generation, application of physics-induced machine learning models or rather from the organization point of view: sharing the data across companies' borders.

If insufficient data are available, ML models tend not to generalize well and there is an increasing risk of overfitting. In DL, one technique that mitigates that risk is data augmentation~\cite{perez2017effectiveness}, which has been shown to improve the performance of DL models in image classification. Recently, also automatic approaches for data augmentation have been proposed, such as AutoAugment~\cite{cubuk2019autoaugment}. However, research for data augmentation on time series data has been limited. An investigation into how data augmentation can be applied in PHM contexts, particularly for time series data, is one of the potential research directions.

Recently, several approaches have been proposed to generate faulty samples or fault features with generative neural networks ~\cite{yu2019cwgan, zhou2020deep}. However, most of the studies focused on vibration data and signals that were pre-processed to make them image-like. An interesting research direction is to evaluate the transferability of such approaches to more complex datasets and time-series data. A further evaluation of the physical plausibility of the generated samples and the effects of the different generated faults on the performance of the algorithms is also required.

An additional challenge that needs to be addressed in future research is effective and efficient composition and selection of the training datasets. This is particularly relevant in evolving environments with highly varying operating conditions where the training dataset is not fully representative of the full range of the expected operating conditions. A decision needs to be taken continuously if the newly measured data should be included in the training dataset and the algorithms should be updated or if the information is redundant and already contained in the dataset used for training the algorithms. This becomes even more complex if the decision on including the newly measured data in the training dataset is taken on the fleet level and not only on the single unit level. The research also goes in the direction of active learning: selecting the observations that will have the largest improvement on the algorithm's performance. 

Uncertainty propagation should also be addressed more comprehensively, in particular by developing or enhancing  Monte-Carlo simulation methods that guarantee sufficient  accuracy with reasonable computation load in presence of nonlinear, non-Gaussian, non-stationary stochastic processes. Synergies with traditional reliability engineering techniques may be exploited more effectively. 

DL provides many promising directions in PHM with the potential to improve the availability, safety and cost efficiency of complex industrial assets. However, there are also several requirements for the industrial stakeholders that need to be fulfilled, before a significant progress can be realised. These requirements include automation and standardization of data collection, including particularly maintenance and inspection reports and implementation of data sharing across several stakeholders; as well as perhaps generally accepted ways of assessing data quality.  An additional hurdle for a broad implementation of DL in PHM is one of the general  hurdles in PHM: development of suitable business models and underlying legal frameworks that foster collaboration between stakeholders and provide benefits for all the involved partners.

\section*{Acknowledgment}
The contributions of Olga Fink and Qin Wang were funded by the Swiss National Science Foundation (SNSF) Grant no. PP00P2\_176878.

\bibliographystyle{./elsarticle-num}
\bibliography{jfms}

%





\end{document}